# Evolution of microstructural and mechanical properties of nanocrystalline $Co_2FeAl$ Heusler alloy prepared by mechanical alloying

M.Hakimi[1,*], P.Kameli[1], H.Salamati[1], Y.Mazaheri[2]

1 Department of Physics, Isfahan University of Technology, Isfahan 84156-83111, Iran

2 Department of Materials Engineering, Isfahan University of Technology, Isfahan 84156-83111, Iran

**Abstract**

Mechanical alloying (MA) has been used to fabricate the $Co_2FeAl$ Heusler alloy with a nanocrystalline structure. The formation mechanism of the alloy has been investigated. Rietveld analysis showed that all samples that were milled for more than 15 hours had an $L2_1$ structure with a space group of Fm3m. The crystallite size and internal strain of the samples were calculated using the Williamson–Hall equation. With mechanical alloying of up to 20 hours the crystallite size of $Co_2FeAl$ increased, after which the crystallite size started to decrease. In contrast, internal strain first decreased during the process and then increased with the increase of milling time. The powder obtained after 20 hours of MA was split into three parts and separately annealed at 300, 500 and 700 $^oC$ for 5 hours. A considerable increase was observed in the hardness value of powder particles with the increase of annealing temperature up to 500 $^oC$. However, the hardness value of the sample annealed at 700 $^oC$ decreased. It seems that this feature is related to parameters such as increase of crystallite size, enhancement of lattice ordering, change in density of defects and impurities and nonstoichiometric effects.

*Corresponding author. Tel: +98 3113912375, fax: +98 3113912376, E-mail: m_hakimi@ph.iut.ac.ir

# 1. Introduction

Electronics is based on charge degrees of freedom. Recently, electronic devices exploiting the spin of electrons, i.e. spintronics, have attracted great scientific attention. Some basic problems in spintronics include producing, transferring and characterizing the spin polarized electrical current. Half-metallic ferromagnets (HMFs) seem to be a good candidate for solving these problems. To date, four types of HMFs have been proposed by calculations, specifically oxide compounds [1], perovskites [2], zinc-blende compounds [3] and Heusler alloys [4]. Some Heusler alloys have been theoretically predicted to be HMFs with a unity spin polarization [5]. Co-based Heusler compounds are of particular interest because they are good ferromagnets, show comparatively high Curie temperatures and have low degrees of atomic disordering [6]. Recently, Galanakis studied the $Co_2CrAl$ full-Heusler quaternary alloy and theoretically predicted a spin polarization of 84% [7]. Kelekar et al. [8, 9] experimentally studied $Co_2Fe_xCr_{1-x}Al$ films that had been grown on MgO substrate. Their findings indicated an enhancement in atomic ordering with increase of x values. Also, the spin polarization of $Co_2Fe_xCr_{1-x}Al$ at $E_F$ increased with the increase of Fe concentration. In addition to high spin polarization, Heusler compounds have recently been investigated due to their magnetoresistance properties [10], exchange bias [11], and thermoelectric [12] and shape memory effects [13]. These properties strongly affect and are affected by the mechanical properties of the compound. For example, strain induced by a magnetic field in a lattice structure is affected by shape memory effects [14]. Also, magnetoresistance varies with crystallite size [15]. In some materials, magnetic properties are affected by lattice parameters [16, 17]. Since the synthesis method and respective processes in sample preparation affect mechanical properties, investigation of preparation processes may be of great value. Polycrystalline Heusler samples are usually synthesized by arc-melting elements in a noble gas atmosphere. For example, Wurmehl et al. [18] investigated the structural properties of $Co_2Cr_{1-x}Fe_xAl$ particles fabricated by arc-melting. Umetsu et al. [19] have studied the magnetic properties of the $Co_2MnAl$ alloy obtained by induction melting in an argon atmosphere. The limitation on amount of product and the generation of inhomogeneous materials due to high speed of production are related to difficulties in arc-melting. The mechanical alloying (MA) method has been recognized as one of the most powerful methods for the production of novel, high-performance, low-cost materials such as ferrites, oxides, intermetallic materials and Heusler alloys [20–25]. For example, $Fe_{1-x}Co_x$, Ni–Fe–Co and $Ni_{0.77}Fe_{0.16}Cu_{0.05}Cr_{0.02}$ have been prepared by MA [26-28].

Furthermore, Mousavi et al. [29] have synthesized NiTi by MA and studied the phase preparation steps. There are few reports on synthesis of Heusler alloys by MA. Recently, Hatchard et al. [22] have explored an MA technique for processing Ni–Mn–Ga Heusler alloys. Vinesh et al. [23] investigated the effect of ball milling on magnetic properties of $Fe_2MnAl$ Heusler alloys. Mikami et al. [25] have also used MA for the preparation of $Fe_2VAl$ Heusler alloys. In our previous work [30], $Co_2CrAl$ Heusler alloy was prepared by MA and the magnetic properties of the samples were investigated. Mechanical properties were strongly affected by the synthesis processes followed in the MA method. For example, in one study, the crystallite size of NiTi decreased rapidly with the increase of MA time; in contrast, lattice distortion increased with longer MA times. Moreover, the annealing process carried out after MA led to the formation of new phases [29]. In the present study, the preparation process of nanocrystalline $Co_2FeAl$ alloy powders prepared by the MA method has been investigated. Since the physical features of Heusler alloys are influenced by structural and mechanical properties, the effects of milling time and annealing temperature on the structural and mechanical properties of this compound have been investigated.

**2. Experiment**

High purity Co, Fe and Al powders were used as raw materials. Fig. 1 shows the morphology of the initial powders. As can be seen, these three elemental materials are distinguishable from one another. Al powder has a flake shape with an approximate particle size of 10 micrometers and Fe powders have a spherical shape with a particle size in the micrometer range. Agglomerated Co powders possess a submicron particle size. The elemental powders with stoichiometric composition of $Co_2FeAl$ were mechanically alloyed in a planetary ball mill under an argon atmosphere. The experiments were carried out at room temperature using a planetary ball mill (specifically, a Fritsch pulverisette P6) using 20mm diameter hardened steel balls and 500 ml stainless steel containers. The ball-to-powder weight ratio and the rotational speed were 10:1 and 300 rpm, respectively.

To prevent overheating of the containers, the experiments were carried out by alternating between 60 minutes of milling and 10 minutes of rest. Following predetermined intervals of time (0, 1, 2, 5, 10, 15, 20, 30 and 40 hours), samples of about 50 mg were taken for X-ray diffraction (XRD) investigations. XRD measurements were performed using a Philips X'PERT MPD diffractometer (Cu Kα radiation: $\lambda$ = 0.154 nm at 20 kV and 30 mA). Crystallite size and lattice distortion were calculated using the Williamson–Hall equation. Rietveld fitting was utilized to define the lattice parameters and space group of the samples. The microstructural observations of samples were analyzed through a Philips XL 30 scanning electron microscope (SEM). In order to ascertain the effect of annealing temperature on the structural and mechanical properties of ball milled powders, three parts of 20h ball milled powders were subsequently annealed under an argon flow at 300, 500 and 700 °C for a period of 5h. The hardness of annealed powders was also determined through microhardness measurements using a Vickers indentor at a load of 500g and dwell time of 10s. For measurement of microhardness, a small amount of powder particles was mounted. Prior to indentation, the surfaces of samples were polished using a series of sandpaper with increasing grits followed by application of a series of diamond pastes.

## 3. Results and discussion

Fig. 2 shows the XRD patterns at room temperature for the mixture of Co, Fe and Al before MA. As expected, the XRD pattern of the unmilled mixture contained only the individual reflections of Co, Fe and Al phases. A noticeable point is that Co appeared in the mixture of both fcc and hcp phases. The intensity ratio of individual reflections was in accordance with the stoichiometric composition of the mixture.

To investigate the formation mechanism of $Co_2FeAl$, the XRD patterns for the compounds with milling times less than 15h have been measured and presented in Fig. (3.a). Fig. 3 also shows the behavior of the diffraction lines of Co and Al by milling time. As can be seen, the intensity of diffraction peaks of Al and Co (fcc-structure) quickly decreased with increase of MA time (Fig. 3.b, 3.d) to the extent that after two hours of milling, these peaks disappeared. This behavior has also been observed in the work of Golubkova et al. [31]. They report that CoAl was formed after one hour of MA. Analysis of x-ray data confirmed the existence of the CoAl phase in the sample obtained after 2 hours of MA. However, the diffraction peaks of Co (hcp-structure) remained in the pattern up to 10 hours of MA (Fig. 3.c). Reports related to preparation of CoFe and FeAl binary compounds show that this behavior is not unexpected [26, 32]. For example, CoFe samples have been obtained after 12 hours of MA [26]. The rate of reaction between Fe and Al is lower than that of Fe and Co. Accordingly, the formation of FeAl is reported to require 100 hours of MA [32]. In another work, Azizi et al. [33] have reported the formation of $Fe_{50}Co_{50}$ to have taken 30 minutes of MA.

The XRD patterns of the 15 h ball-milled powder seem similar to those of an $L2_1$-$Co_2FeAl$ Heusler phase. Additionally, there is no noticeable trace of starting materials in the XRD patterns of the powder. Also, Rietveld analysis showed that the compounds with milling times of over 10 hours were single phase and had an $L2_1$ structure with space group of Fm3m.

For example, the Rietveld refinement pattern of the 40 h milled sample is presented in Fig. 4. Experimental data (red line), theoretical data (black line), the difference between experimental and theoretical data (blue line) and Bragg-positions (green line) are shown in the figure.

The lattice parameters obtained by Rietveld fitting are shown in Table 1. There was no noticeable change in the estimated lattice constants with increase of milling time. The lattice constants conform with the work done by Wurmehl [18].

Fig. 5 shows the XRD patterns at room temperature for the $Co_2FeAl$ powder compound at different milling times. As mentioned earlier, with the increase of milling time, the diffraction lines of the initial powders disappeared and the Fm3m phase emerged. Furthermore, the diffraction peaks of the $Co_2FeAl$ phase heightened with the increase of milling time up to 15 hours. This is characteristic of the formation of this phase. For the samples that were ball milled for more than 15h, a broadening and a decrease in peak height were observed. A main cause for this behavior may be the increase of lattice strain and decrease of crystallite size, leading to an enhancement of surface effects. The crystallite size and lattice strain of the $L2_1$-$Co_2FeAl$ Heusler phase were calculated using the Williamson–Hall equation. The Williamson–Hall equation is expressed in the following manner [34–36]:

$$B\cos(\theta) = 2(\varepsilon)\sin(\theta) + k\lambda/D$$

Where B is the full width at half maximum (FWHM) of the XRD peaks, K the Scherer constant, D the particle size, $\lambda$ the wavelength of the X-ray, $\varepsilon$ the lattice strain and $\theta$ the Bragg angle.

The crystallite size and lattice strain of the samples are shown in Fig. 6 as a function of MA time. The crystallite size of $Co_2FeAl$ increased to about 22 nm after 20 h of MA and then decreased with further milling. The initial increase of crystallite size may be caused by formation of the Heusler phase. Also, the breaking of crystallites during the milling process led to a decrease in crystallite size. Some factors such as crystal point defects, surface defects and dislocations can introduce internal strain to the lattice. The lattice strain increased with further MA. In other words, this effect may be due to increase in the volume of dislocations with the decrease of crystallite size. Similar results have been reported in other systems as well [29].

Fig.7 shows typical SEM images for zero, 1 and 15 hour milled samples. The 1h milled sample shows a relatively broad distribution of agglomerates composed of strongly welded nanocrystallites of inhomogeneous morphologies with particle sizes in the range of about 50 micrometers (7.b). In the first hours of milling, the particles clung to each other because of the collision of the balls and powder particles. For this reason, surface contact of the particles increased. This agglomeration provides a suitable condition for the solution of materials. Further milling led to the formation of the final phase that was accompanied by breaking of the particles. In the 15h milled sample, narrow distribution of agglomerates were observed with a particle size in a range of about 10 micrometers (7.c).

In order to study the effect of annealing on structural properties of the samples, the 20 h mechanical alloyed powder was annealed under an argon flow at 300, 500 and 700 $^{o}$C for 5 hours. Fig. 8 shows the XRD patterns at room temperature for the samples. It is evident that the intensity of XRD peaks increased after annealing due to an increase in crystallite size as well as elimination of internal strain. The crystallite size and lattice strain of the annealed samples are shown in Fig. 9 as a function of annealing temperature. The crystallite size of $Co_2FeAl$ increased and internal strain in the lattice decreased with the increase of annealing temperature. Heating processes can provide the possibility of an interaction between crystallite boundaries leading to the formation of bigger size crystallites. Also, using a heating process to intensify lattice oscillations can facilitate elimination of lattice dislocations and therefore decrease internal strain.

Fig. 10 shows the average microhardness values of powder particles at different annealing temperatures. A considerable increase was observed in hardness values of powder particles with the increase of annealing temperature up to 500 $^{o}$C. However, the hardness value of the sample annealed at 700 $^{o}$C decreased. It seems that this feature is related to parameters such as increase of crystallite size, enhancement of lattice ordering, changes in density of defects and impurities and nonstoichiometric effects [29, 37].

## 4. Conclusion

The $Co_2FeAl$ Heusler alloy was prepared by mechanical alloying and the synthesis process was investigated step by step. It seems that the formation of the Heusler phase is an appropriate explanation for the increase of crystallite size in the first hours of MA. Also, the breaking of grains during the milling process led to a decrease in grain size for the samples with further milling. Furthermore, increase in lattice strain was caused by increase in the volume of dislocations with the decrease of crystallite size through MA. According to typical SEM images during the first hours of milling, agglomeration of initial powders provided a suitable condition for the solution of materials. Further milling led to formation of the final phase that was accompanied by breaking of the particles to the extent that in the 15h milled sample, narrow distribution of agglomerates could be observed with a particle size in the range of about 10 micrometers. During the annealing process, the crystallite size increased and internal strain decreased. Also, in the annealing process the average microhardness values of powder particles increased at first and then decreased at 700 $^{o}$C.

**Acknowledgments**

The authors would like to thank Isfahan University of Technology for supporting this project.

Figure captions:

Fig.1. SEM morphology of initial powders

Fig. 2. The XRD patterns for the mixture of Co, Fe and Al before MA

Fig. 3. a) The XRD patterns for the compounds with milling time less than 15h. b, c and d) The behavior of the diffraction peaks of Co and Al by milling time

Fig. 4. Rietveld refinement pattern of 40 h milled sample

Fig. 5. XRD patterns of the $Co_2FeAl$ powder compound at different milling times

Fig.6. Crystallite size and lattice strain of the samples as a function of MA time

Fig. 7. Typical SEM images for unmilled, 1 and 15h milled samples

Fig. 8. XRD patterns for the samples annealed at 300, 500 and 700 $^oC$

Fig. 9. Crystallite size and lattice strain of the annealed samples as a function of annealing temperature

Fig. 10. Average microhardness values of powder particles at different annealing temperatures

Table.1. Lattice parameters of samples

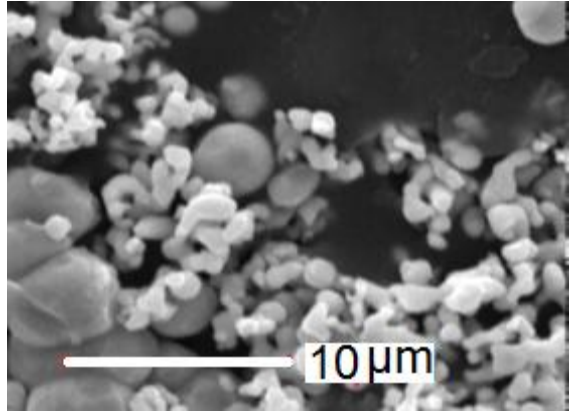

Fig.1

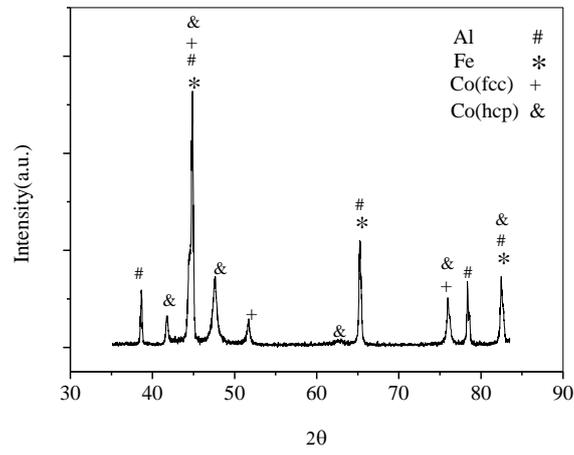

Fig. 2

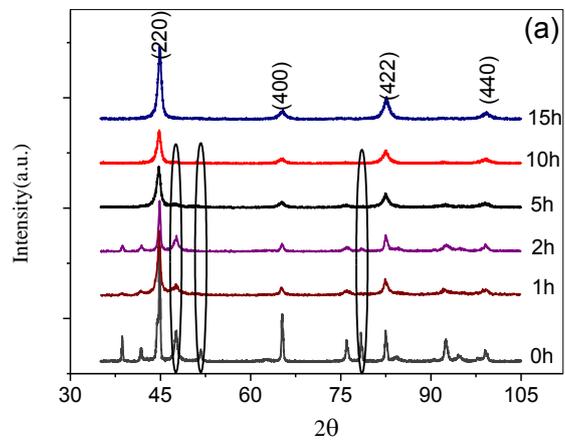

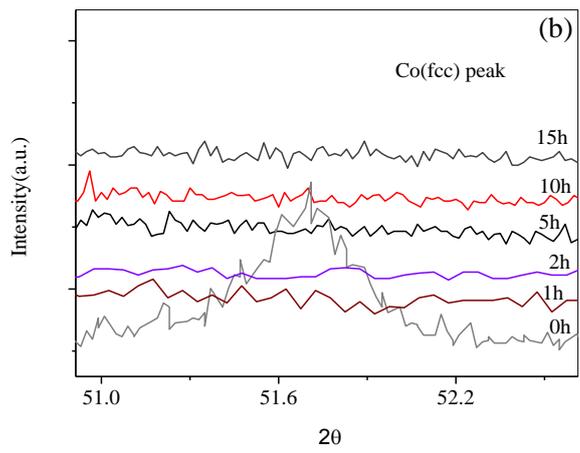

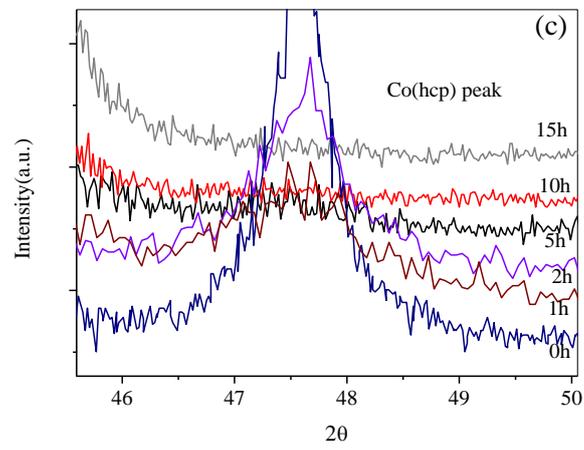

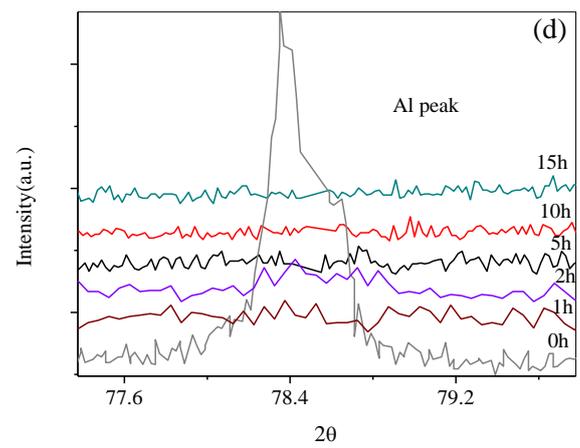

Fig. 3

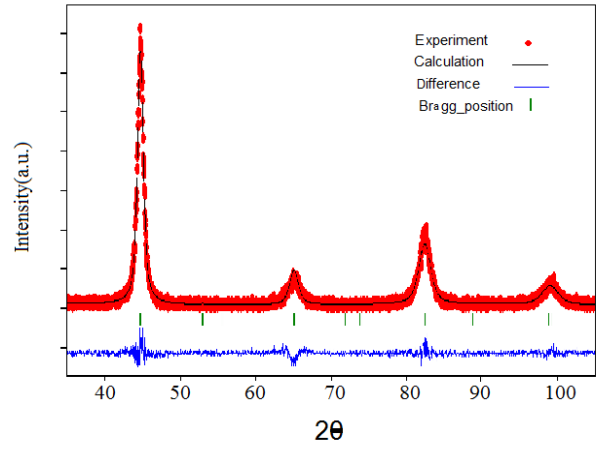

Fig. 4.

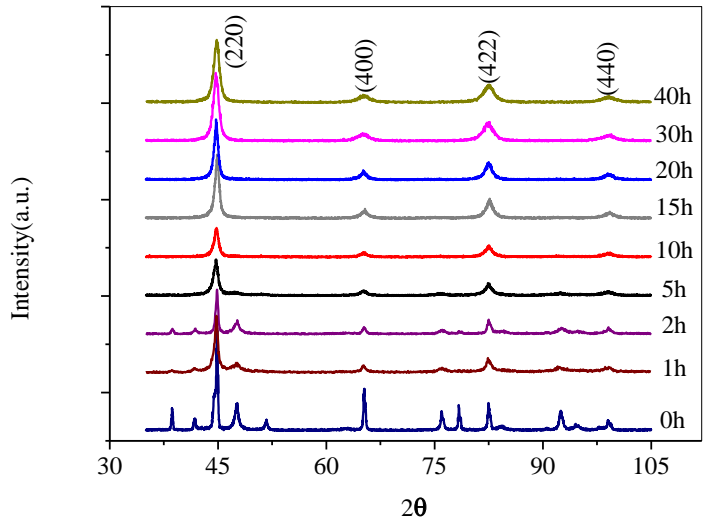

Fig. 5

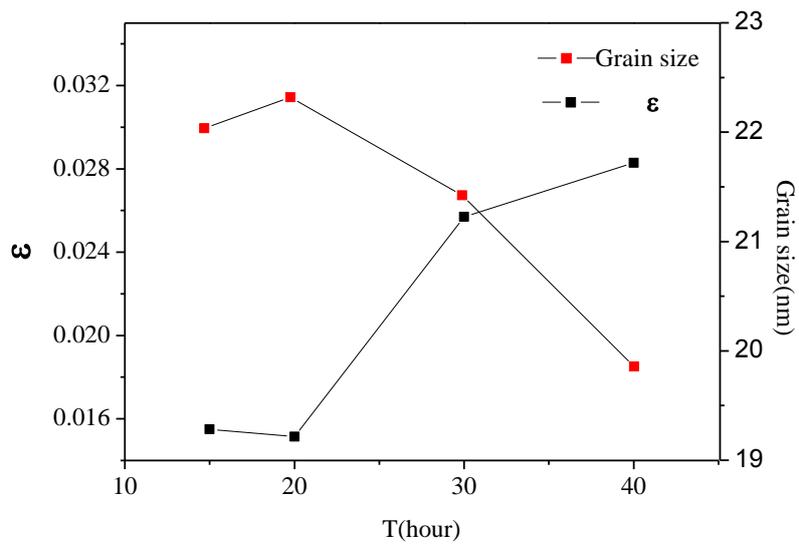

Fig.6

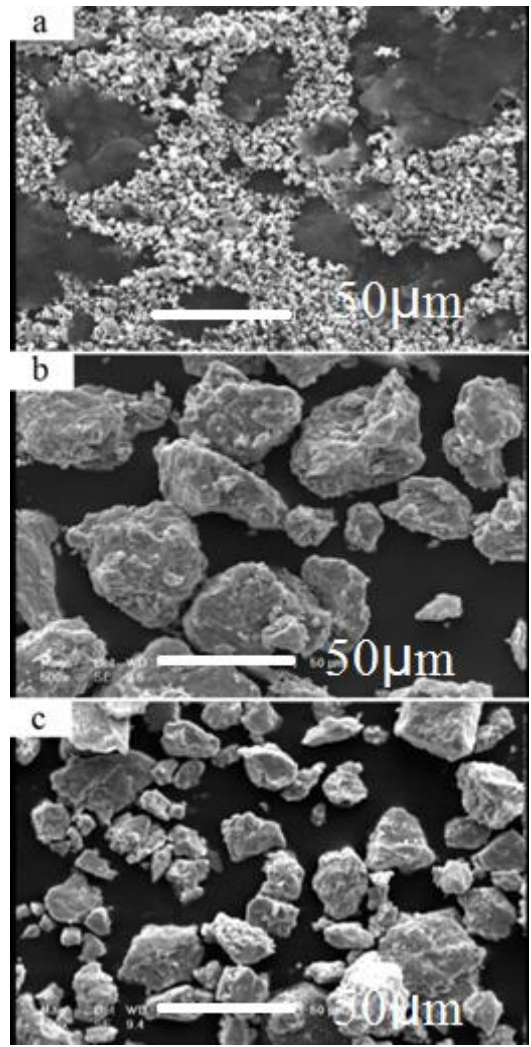

Fig. 7

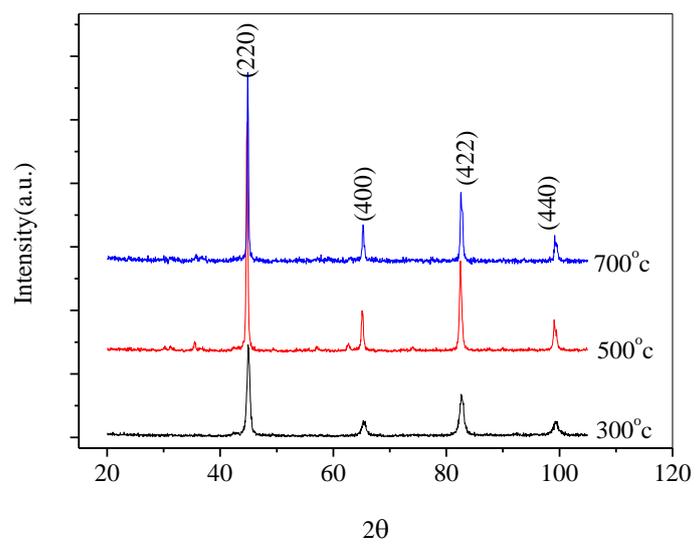

Fig. 8

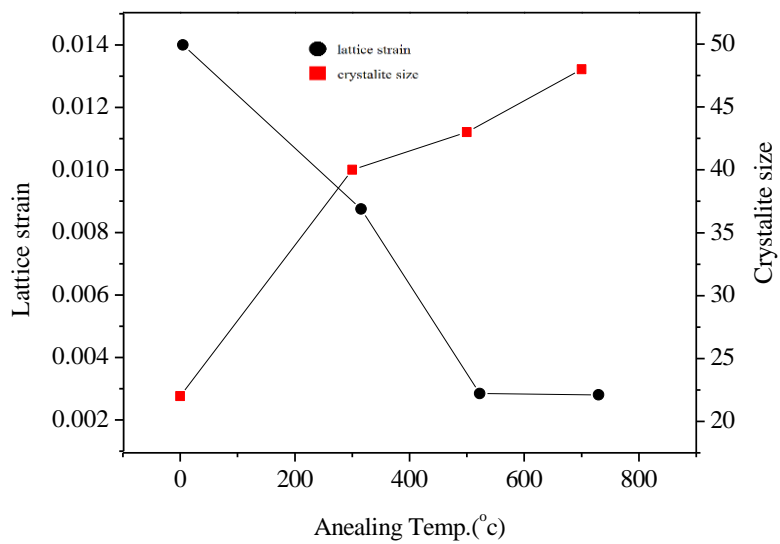

Fig. 9

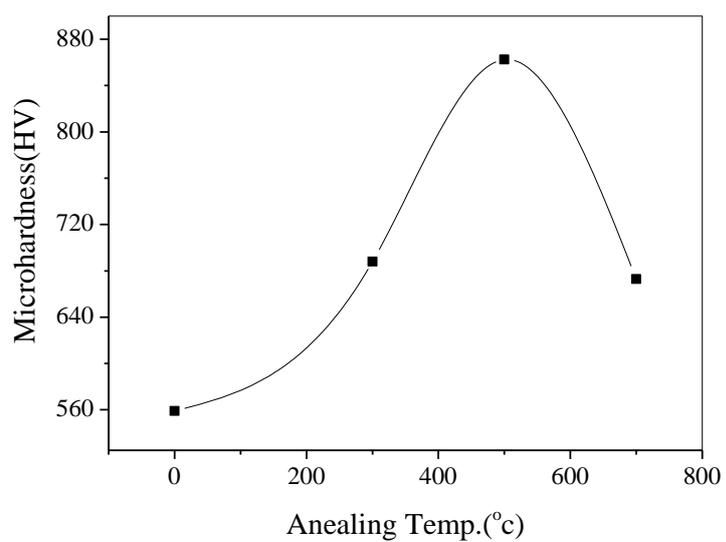

Fig. 10

| Milling time (h) | Lattice parameter (Å) |
|---|---|
| 15 | 5.73(1) |
| 20 | 5.72(8) |
| 30 | 5.73(8) |
| 40 | 5.73(7) |

Table.1